\documentclass{article}
\usepackage{spconf,amsmath,graphicx,hyperref, bm}
\usepackage{amssymb}
\usepackage{booktabs}
\usepackage{multirow, scalerel}
\usepackage{tabularray, tabularx}

\usepackage{fancyhdr}
\fancypagestyle{firstpage}{
  \fancyhf{}

  \fancyfoot[C]{\scriptsize Jayeon Yi and Minje Kim, ``From Hallucination to Articulation: Language Model-Driven Losses for Ultra Low-Bitrate Neural Speech Coding,"\\
  In Proc. of the IEEE International Conference on Acoustics, Speech and Signal Processing (ICASSP), Barcelona, Spain, 2026.}
}
\fancypagestyle{plain}{
  \fancyhf{}

}
\newlength{\myMheight}
\title{From Hallucination to Articulation: Language Model-Driven Losses for Ultra Low-Bitrate Neural Speech Coding}
\name{Jayeon Yi and Minje Kim\thanks{This work was supported by Electronics and Telecommunications Research Institute (ETRI) grant funded by the Korean government [26ZC1100, Development of Spatial Media Technology and Interaction Technology for Convergence of the Real and Virtual World].}}
\address{University of Illinois Urbana-Champaign, Siebel School of Computing and Data Science, USA 61801}
\begin{document}
\ninept
\maketitle
\begin{abstract}
``Phoneme Hallucinations (PH)'' commonly occur in low-bitrate DNN-based codecs. It is the generative decoder's attempt to synthesize plausible outputs from excessively compressed tokens missing some semantic information. In this work, we propose language model-driven losses (LM loss) and show they may alleviate PHs better than a semantic distillation (SD) objective in very-low-bitrate settings.
The proposed LM losses build upon language models pretrained to associate speech with text. When ground-truth transcripts are unavailable, we propose to modify a popular automatic speech recognition (ASR) model, Whisper, to compare the decoded utterance against the ASR-inferred transcriptions of the input speech. Else, we propose to use the timed-text regularizer (TTR) to compare WavLM representations of the decoded utterance against BERT representations of the ground-truth transcriptions. We test and compare LM losses against an SD objective, using a reference codec whose three-stage training regimen was designed after several popular codecs. Subjective and objective evaluations conclude that LM losses may provide stronger guidance to extract semantic information from self-supervised speech representations, boosting human-perceived semantic adherence while preserving overall output quality. Demo samples, code, and checkpoints are available online \footnote{https://minjekim.com/research-projects/lm-loss\#icassp2026}.
\end{abstract}
\section{Introduction}
\label{sec:intro}


Deep neural networks (DNNs) have emerged as a viable model for low-bitrate speech codecs. While traditional codecs such as 
MELPe \cite{melpe} or Codec2 \cite{codec2} are known to enable intelligible transmissions at sub-1 kbps rates at the cost of reduced sound quality, many DNN-based codecs have successfully reached such ultra-low bitrates with improved intelligibility and perceptual sound quality \cite{soundstream, encodec, dac, focalcodec, semanticodec}. 

Hence, the training objectives for DNN-based speech codecs are to learn to convert speech to a compact (i.e., low-bitrate) discrete representation and then decode it back to audio with the least possible difference from the input. More specifically, there are two major categories of codecs based on the training procedures and input features, as commonly classified \cite{focalcodec, repcodec, speechtokenizer}. First, \textit{acoustic} codecs rely more on reconstruction losses defined within a short period of signals \cite{soundstream, encodec, dac}. 
Conversely, \emph{semantic} codecs aim to preserve semantic information during the coding process, often by learning from foundation models trained with longer-term self-supervision objectives, such as HuBERT \cite{hubert} or WavLM \cite{wavlm}, which are thought to contain rich semantic information in their output. Many successful state-of-the-art codecs \cite{focalcodec, semanticodec, stablecodec, xcodec, xy_tokenizer} are of this category. 

In this paper, we focus on a specific type of coding artifact, ``phoneme hallucination (PH)," which results from generative decoders' attempts to synthesize plausible outputs from excessively compressed codes that may lack some essential semantic information.
As a remedy, many codecs propose to 
adapt semantic distillation (SD) procedures \cite{speechtokenizer, xcodec, moshi}, or use self-supervised representations as inputs \cite{focalcodec, semanticodec, that_one_google_paper, reference_primo}.
Indeed, semantic losses seem to emphasize linguistic accuracy in the reconstruction.

However, semantic codecs such as TAAE \cite{stablecodec}, FocalCodec \cite{focalcodec}, and SemantiCodec \cite{semanticodec} do exhibit PHs at low bitrates ( $<0.4$ kbps), as made evident in listening examples shared online by the respective authors. 
On the other hand, it has been shown possible to revert self-supervised speech representations to speech \cite{nansy}. This implies that better distillation objectives may exist, with the potential to benefit existing and new architectures alike.


Meanwhile, automatic speech recognition (ASR) models, such as Whisper \cite{whisper}, are trained to directly estimate subwords 
given a speech utterance, where the model's ability in learning the long-term linguistic context is critical to its success. 
Similarly, the timed-text regularizer (TTR) \cite{tsunan} directly maps WavLM \cite{wavlm} representations to those learned by the corresponding language model, BERT \cite{bert}, to introduce linguistic contexts to speech enhancement results. 
Since these models were explicitly trained to associate speech representations with text representations, they may be a richer source for ultra-low bitrate speech codes to learn the semantics from.

Some previous works in speech coding do use such linguistic context to enhance token quality.
For instance, LLM-Codec \cite{llm_codec} proposes to fill and freeze its codebooks with large language model (LLM) embeddings for thousands of carefully selected words, along with an L2 loss enforcing similarity between the VQ outputs and LLM embeddings. However, the bitrate is not as competitive as other semantic codecs. 
As another example, XY-tokenizer \cite{xy_tokenizer} distills from an LLM to map speech tokens to LLM input so that the LLM would output a transcription for the speech. In this way, the learned tokens embed semantic information transferred from the LLM. Again, XY-tokenizer operates in a much higher bitrate (1 kbps). In summary, while LLMs see some recent usage as a part of distillation objectives for codecs, proposed architectures in the literature 
introduce additional bits to host the semantic information.


Meanwhile, there are also non-speech-codec works, e.g., \cite{that_one_se_paper}, that attaches adapters \cite{adapter} to a speech enhancement model to distill from an LLM. However, adapters induce additional inference overhead, while also blocking the loss flow. In contrast, \cite{tsunan} uses their TTR model as a regularization loss, inducing zero inference overhead while still improving their speech separation model. However, the work does not examine whether their scheme improves the linguistic conformance of output speech. Neither of these approaches has been evaluated for training codecs, either.

In this work, we formulate and test \emph{language model-driven losses} (LM loss), demonstrating how they can enhance a neural speech codec. We first show that PHs can still manifest in a simple semantic codec trained on widely used training procedures. Then, we formulate two families of LM losses. While these do leverage pretrained models like some of the previous works, our proposed methods do not levy any architectural constraints or require any additional finetuning once pretrained; they can be applied to any model that outputs speech. We show that many tested variants of LM losses result in suppressed PHs. Finally, we compare the strengths and drawbacks of each approach with objective metrics, including word error rate (WER), as well as a MUSHRA-like subjective test and an intelligibility test.  





\section{LM Losses for Codecs} 
\label{sec:linguisticdistillation}




\subsection{The Phoneme Hallucination (PH) Phenomenon} 
\label{sec:phoneme_hallucination} 


Although a past work finds $\sim$100bps to be a minimum bitrate for speech communication \cite{theoretic}, many codecs often operate with a higher bitrate due to their inefficiency and the need for sending extensive acoustic information on top of the semantic contents. In this work, we focus on a specific type of artifact a modern neural codec can generate, namely \textit{phoneme hallucination} (PH), where the codec reconstructs incorrect phonemes. It is a failure mode related more to the semantic aspect of the utterance than acoustics, as a PH can sound clean. We observe PHs when an aggressive reduction of bitrate is coupled with generative decoding, as in very-low-bitrate codecs like \cite{focalcodec, semanticodec, stablecodec}. For example, the encoder's temporal decimation could be too much, to the level that the frame rate of the code is too low to represent a phoneme. Having too few codewords in a codebook could also lead to misrepresentation: instead of producing acoustic noise, a generative decoder can try to come up with a plausible phoneme different from the original one. We call it hallucination because the synthesized phoneme could still sound clean, meaning we need an advanced semantic loss function that measures this discrepancy to tame the behavior of the generative decoder, while potentially injecting more information to the low-bitrate code.
\begin{figure}[t]
  \centering
  \includegraphics[width=\linewidth]{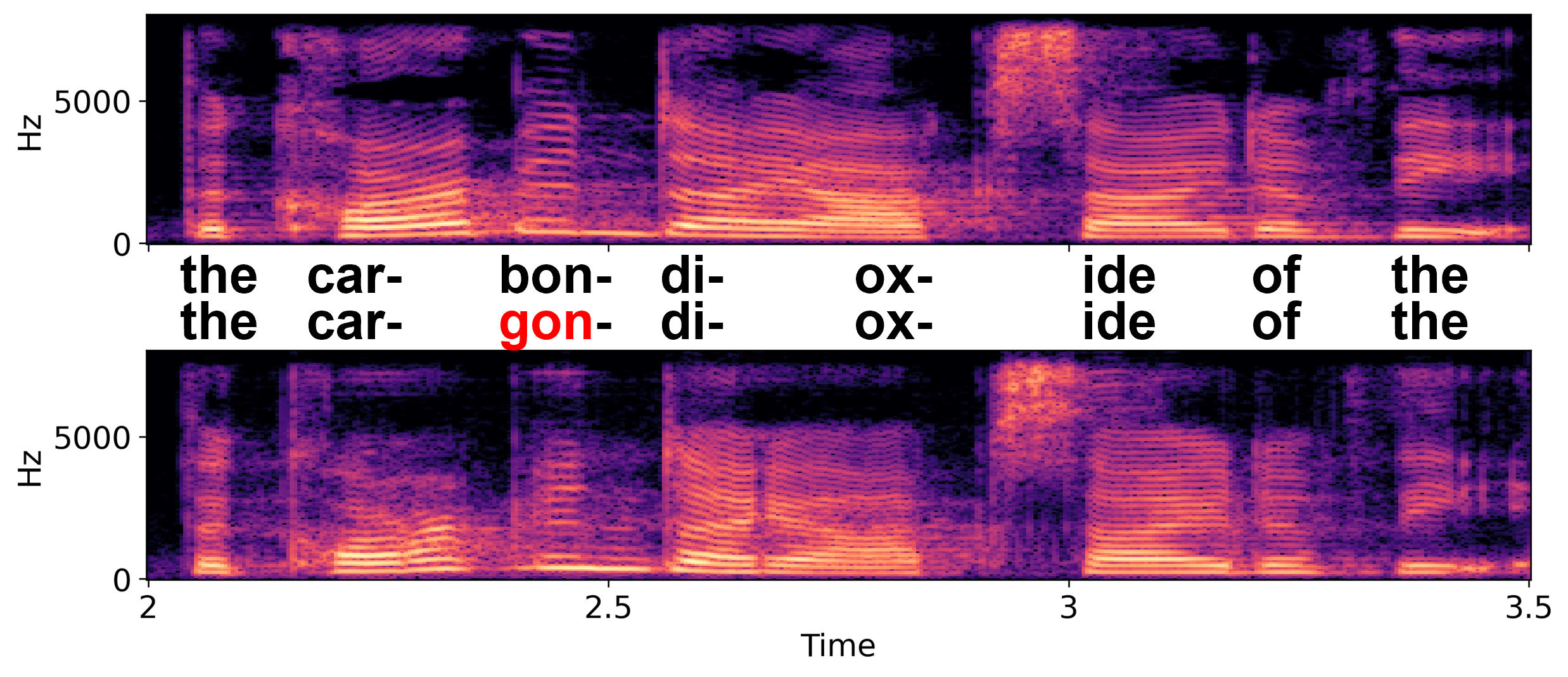}
  \label{fig:phoneme_hallucination}
  \vspace{-0.3in}
  \caption{ (Top) The input speech. (Bottom) The 187.5 bps reference codec with no LM losses exhibits phoneme hallucinations (PHs).}
\end{figure}

\subsection{LM Losses}
\label{sec:linguistic_losses}

The proposed LM losses distill linguistic information from pretrained LMs to explicitly match speech and text. They can be applied to any DNN-based speech codec with no architectural modifications or additional inference overhead.
Generally, an LM loss takes the decoded utterance as input and measures its linguistic feasibility by comparing the two semantics extracted from the decoded utterance and a corresponding script. We propose two loss functions to cover both cases: with and without ground-truth transcripts. 

\subsubsection{The automatic speech recognition (ASR) loss}
The ASR loss leverages ASR models pretrained on a subword-wise autoregressive transcription task as in \cite{whisper}. Let an ASR model $\mathcal{S}$ be trained to predict the next subword $\bm{w}_{i+1}$ in the learned token space, 
\begin{equation}\label{eq:asr_clean}
\bm{w}_{i+1}\approx\hat{\bm{w}}_{i+1}\leftarrow \mathcal{S}(\bm{x}, \widehat{\bm{W}}_{<i+1}),
\end{equation}
given a speech utterance ${\bm{x}}$ and all the previously predicted subword tokens $\widehat{\bm{W}}_{<i+1}:=[\hat{\bm w}_1, \hat{\bm w}_2, \ldots, \hat{\bm w}_i]$. 

We repurpose this autoregressive loss function to measure the fitness of the codec's decoded signal $\hat{\bm{x}}$. To this end, we first predict the token sequence $\widehat{\bm{W}}$ by using the clean speech $\bm{x}$ as the input audio as in \eqref{eq:asr_clean}. Then, we predict another token sequence $\widetilde{\bm W}$ by using the decoded signal $\hat{\bm{x}}$ and the token sequence $\widehat{\bm{W}}$ predicted from the clean speech $\bm{x}$,
\begin{equation}\label{eq:asr_loss}
    \mathcal{L}_\text{ASR}(\bm x, \hat{\bm x})=\frac{1}{N}\sum_{i=1}^N \mathcal{L}_\text{CE}\big(\mathcal{S}(\hat{\bm x}, \widehat{\bm W}_{<i})||\hat{\bm w}_i\big),
\end{equation}
where $\mathcal{L}_\text{CE}$ is the subword level prediction loss defined over tokens, e.g., cross entropy.  



The main advantages of the proposed ASR loss are that it can leverage the internal LM that the ASR model already learned from the pairs of the clean utterance $\bm x$ and its corresponding tokens $\bm W$. In addition, the loss defined in the token space does not need a ground-truth script, unleashing the possibility of using any clean speech corpora for codec training. While one can easily modify this loss to use a ground-truth script, we found in preliminary experiments that this may cause significant training instabilities.


\begin{figure}[t]

  \centering
  \includegraphics[width=\columnwidth]{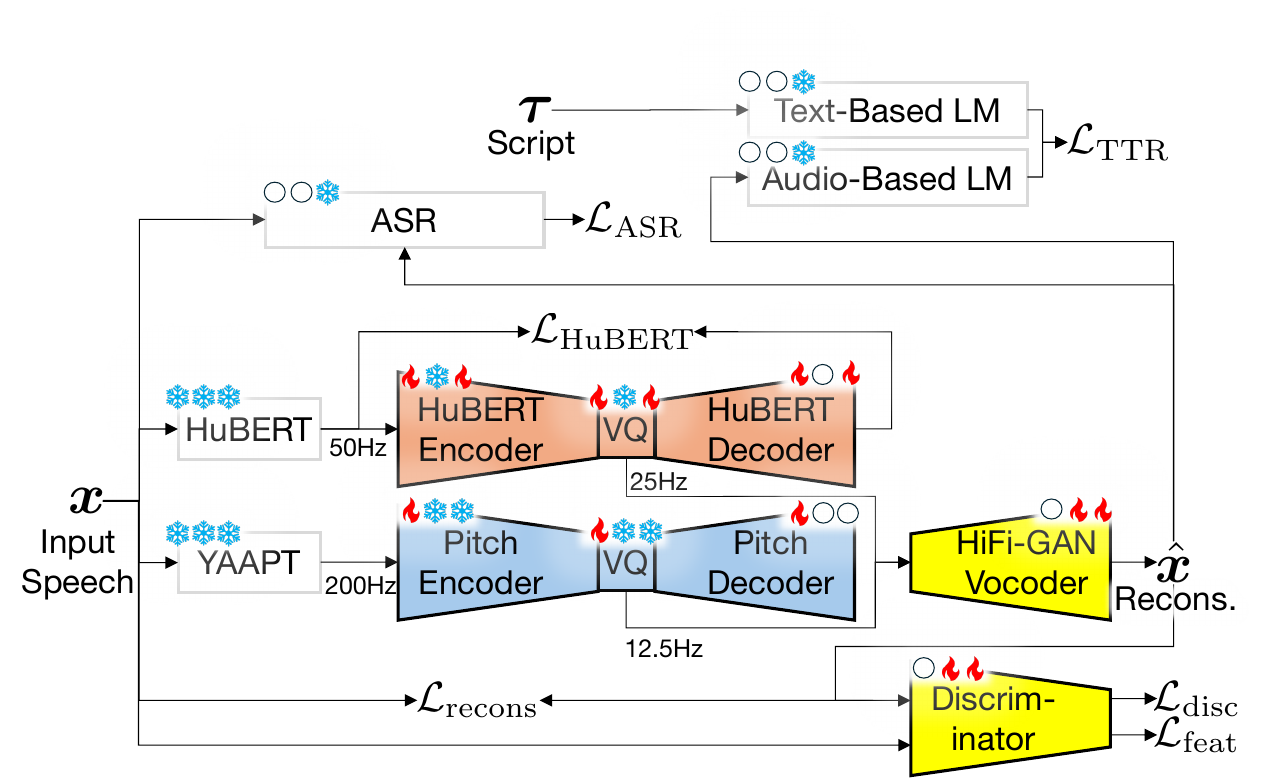}
  \vspace{-0.2in}
  \caption{ Architecture and training of our reference codec. Our three-stage training procedure emulates common codec-training setups. \includegraphics[height=\myMheight]{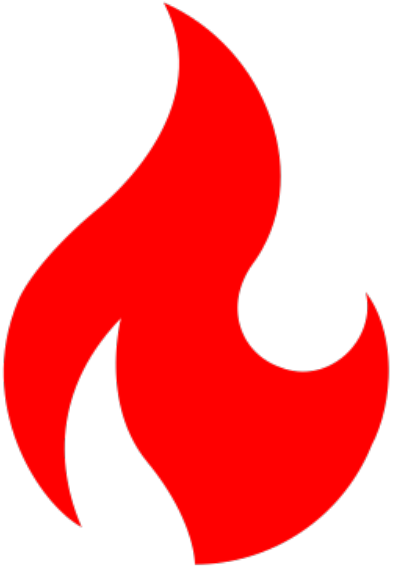} denotes the modules updated in the given stage, while \includegraphics[height=\myMheight]{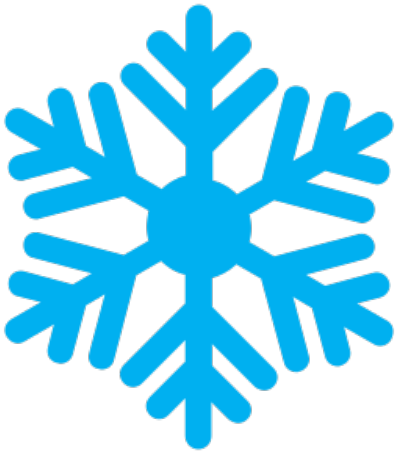} represents frozen ones. \includegraphics[height=\myMheight]{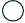} does not participate in. For example, \includegraphics[height=\myMheight]{torchlight.pdf}
\includegraphics[height=\myMheight]{snowflake.pdf}
\includegraphics[height=\myMheight]{null.pdf} means that the module is updated in the first stage, while being frozen for the second stage, and then being idle in the third stage. In the third stage, either a LM loss ($\mathcal{L}_{\text{ASR}}$,$\mathcal{L}_{\text{TTR}}$) or the SD loss ($\mathcal{L}_{\text{HuBERT}}$) is employed in combination with the others.}
  \label{fig:codec}
  
  

\end{figure}

\subsubsection{The timed-text regularizer (TTR) loss}

Another approach to imposing more semantics to codec training is to directly use time-aligned scripts as the target, as proposed in \cite{tsunan} for source separation. In the context of our neural speech codec (see Fig. \ref{fig:codec}), the timed-text regularization (TTR) first processes the decoded speech $\hat{\bm x}$ using the audio-based language model (LM) whose output embeddings are compared with the ones from the text-based LM, assuming their subword-level alignment. 

More specifically, the LMs involved in the definition of the TTR loss are pretrained from a clean utterance and its script, so their output embeddings are guaranteed to be similar for clean speech. For the $i$-th subword and its corresponding segment of the speech signal ${\bm x}^{(i)}$, i.e., ${\bm x}=[{\bm x}^{(1)}; \ldots; {\bm x}^{(N)}]$, WavLM \cite{wavlm}, a self-supervised speech model, transforms the variable-length subword audio signal ${\bm x}^{(i)}$ into a series of embedding vectors ${\bm X}^{(i)}=[{\bm X}_{:,1}^{(i)}, \ldots, {\bm X}_{:,m}^{(i)}]$. Since the number of embeddings $m$ varies by the length of the spoken subword, a transformer-based \textit{summarizer} module turns them into a single representative embedding vector: ${\bm S}_{:,i}\leftarrow \mathcal{P}_\text{Sum.}({\bm X}^{(i)})$. After constructing ${\bm S}$ with $N$ total embeddings, we impose the inter-subword relations into this audio-based representation via another transformer-based \textit{aggregator} module, which transforms the entire sequence ${\bm{S}}$ into a self-attended version $\overline{\bm S}$, i.e., $\overline{\bm S}\leftarrow\mathcal{P}_\text{Agg.}({\bm S})$.

Meanwhile, the ground-truth script $\tau$ is processed by the text-based LM, BERT \cite{bert}, to produce the subword-level embedding sequence $\bm{T}$. Since both BERT and WavLM models are pretrained and frozen, the summarize and aggregator layers can be seen as a trainable projection module that converts the audio embeddings into the space defined by $\bm{T}$. Hence, we define the TTR loss function:

\begin{align}
    \nonumber\mathcal{L}_\text{TTR}(&\overline{\bm S}\|\bm{T})=\frac{1}{N}\sum_{i=1}^N \bigg(1-\frac{\overline{\bm S}_{:,i}\cdot\bm T_{:,i}}{\|\overline{\bm S}_{:,i}\| \|\bm T_{:,i}\|}\bigg) \\
    &+ \frac{2}{N(N+1)}\sum_{1\leq i\leq j \leq N} \|\overline{\bm S}_{:,i}\cdot\overline{\bm S}_{:,j}- \bm T_{:,i} \cdot \bm T_{:,j}\|^2. 
\end{align}
The first term calculates the cosine similarity between each pair of $\overline{\bm S}_{:,i}$ and $\overline{\bm T}_{:,i}$, while the second term scores the similarity of internal pairwise relations between $\bm S$ and $\bm T$ vectors.


While BERT and WavLM models are frozen, $\mathcal{P}_\text{Sum.}$ and $\mathcal{P}_\text{Agg.}$ are updated to minimize this loss for the clean signals. 
Once this pretraining is done, the frozen text-based and audio-based LMs are used to calculate the same TTR loss, but by taking the decoded signal $\hat{\bm x}$ rather than ${\bm x}$ to update the modules in the codec.

\section{Experimental Setup}
To test if LM losses can enhance neural speech codecs that already include semantic quality considerations, we design experimental procedures with common semantic codecs in mind.

\subsection{The Reference Codec} 
\label{sec:reference_codec}

Our reference codec in Fig. \ref{fig:codec} is a slight modification of a previous codec \cite{reference_primo} that trained a HiFi-GAN vocoder \cite{hifigan} to decode from quantized pitch and HuBERT \cite{hubert} features. Starting from this codec, we add a HuBERT encoder that decimates the feature rate by a factor of 2. The new encoder is comprised of a \texttt{conv1d} to reduce input dimensions from 768 to 128, followed by a ResBlock that shares architecture with the pitch encoder ResBlocks in \cite{reference_primo}. 
Then, a HuBERT VQ codebook quantizes the encoder outputs. This codebook has a size of 32 or 64. Thus, the bitrate of the codec is 187.5 bps or 212.5 bps. We further expanded pitch codebook size to 32, keeping the bitrate same while enhancing acoustic reconstruction quality. 

Overall, this codec faithfully emulates some common patterns in the literature: self-supervised models as an ``encoder'' constituent 
\cite{that_one_google_paper, focalcodec, semanticodec, xcodec}, an additional ``compressor'' to further decimate self-supervised representations \cite{focalcodec, repcodec, xcodec}, and separation of acoustic and semantic ``branches'' \cite{semanticodec, xy_tokenizer}. 

\begin{table}[t]{}
 \aboverulesep=0ex
 \belowrulesep=0ex
      \centering
      \resizebox{.9999\columnwidth}{!}{%
      \begin{tabular}{c|c|cc|cc}
         \multicolumn{1}{c}{} & \multicolumn{1}{c|}{} & \multicolumn{2}{c|}{Semantic}& \multicolumn{2}{c}{Acoustic} \\
         \cmidrule(lr){3-4} \cmidrule(lr){5-6}
        \multicolumn{1}{c}{Rate} & LM & \multicolumn{2}{c|}{WER(\%)$\downarrow$}  & \multirow{2}*{PESQ$\uparrow$} & \multicolumn{1}{c}{WARPQ$\uparrow$} \\
        \multicolumn{1}{c}{(bps)}& loss &\multicolumn{1}{c}{Whisper}&\multicolumn{1}{c|}{wav2vec2.0}&&\multicolumn{1}{c}{normalized} \\ 
         \midrule
         \midrule
         \multirow{4}*{187.5} 
         & ASR & \textbf{1.45} & \textbf{4.56} &1.35& 0.289 \\
         \cmidrule{2-6}
         &TTR &  2.34 & 7.13& 1.39& 0.293 \\
         \cmidrule{2-6}
         &SD &  3.33  &11.2&\textbf{1.42}& \textbf{0.295} \\
         \cmidrule{2-6}
         & S2 & 3.04 & 8.82&1.35& 0.283 \\
         \midrule
         \midrule
         \multirow{4}*{212.5}& ASR & \textbf{1.23}  &\textbf{3.63}&1.37& .289 \\
         \cmidrule{2-6}
         &TTR &  1.53 &5.25&1.44&  .293 \\
         \cmidrule{2-6}
         &SD &  2.11 & 7.04& \textbf{1.46}& \textbf{.295} \\
         \cmidrule{2-6}
         & S2 & 2.09 & 6.34&1.36& .289 \\
         \midrule
         $\infty$ & - & 0.95 & 1.74 & 4.64& 1.00 \\ 
         \bottomrule
    \end{tabular}
    }
      \centering
    \caption{Objective metrics for all codecs. ASR, TTR, and SD denote codecs from stage 3, respectively trained using the proposed LM losses ($\mathcal{L}_{\text{ASR}}$, $\mathcal{L}_{\text{TTR}}$) and SD loss ($\mathcal{L}_{\text{HuBERT}}$). S2 denotes the stage-2 codec trained on frozen encoder and VQ. Rate of $\infty$ denotes unencoded utterances.}
    \label{tab:the_table}
\end{table}

\begin{figure}[t]
  \centering
  \includegraphics[width=\linewidth]{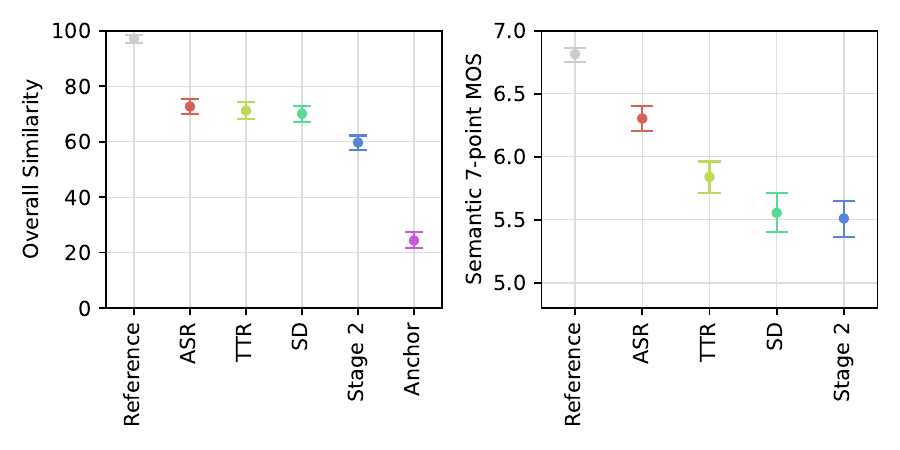}
  \vspace{-0.3in}
  \caption{ Overall similarity (left) and the semantic 7-point MOS (right) subjective evaluations; mean and 95\% confidence intervals. Codecs trained with LM losses show significantly better semantic performance compared to the others.}
  
  \label{fig:subjective_results}
\end{figure}

\subsection{Dataset and Linguistics-Preserving Batching}

To further simplify the acoustic part of the codec, we chose to train on the single-speaker LJSpeech dataset, which could be seen as a specialized module in the personalized neural speech codec as proposed in \cite{JangI2024pnsc}. Combined with the architectural choices and the set of evaluations, this lays out a minimal yet sufficiently fertile framework for evaluating both semantic and acoustic effects of LM losses.

To allow codecs to learn longer semantic contexts, we 
split utterances by their source texts. 
``LJ021'' to ``024`` utterances comprise the test split, while ``025'' to ``027'' comprises the validation split. We also concatenated utterances 
to train on longer (30-45s) segments from the original texts, where each segment starts with a unique sentence. We used a batch size of 1. The pitch and HuBERT embeddings for each utterance were respectively derived using YAAPT \cite{yaapt} and \texttt{HuBERT-base-ls960h} \cite{hubert}. Finally, the text-speech alignments required by TTR were derived with the Montreal Forced Aligner \cite{mfa}.

\subsection{LM loss configurations and TTR modules pretraining}

We use the pretrained \texttt{Whisper-tiny} \cite{whisper} model as the ASR model $\mathcal{S}$. For TTR, we follow \cite{tsunan} to use \texttt{BERT-base-uncased} \cite{bert} and \texttt{WavLM-base} \cite{wavlm}. The summarizer and aggregator are identical four-layer transformer encoder models with a dimension of 768 and a feedforward dimension of 1024.
they are jointly pretrained on the \texttt{LibriSpeech-960h} dataset using similar splitting and batching procedures. The Adam optimizer was used with a learning rate of $1\times 10^{-4}$, $(\beta_1,\beta_2)=(0.8,0.99)$, and with an exponential decay scheduler with $\lambda=0.999$. We use the best-performing checkpoint in terms of validation loss after training for 1M steps.

\subsection{Training the Codec}
\label{sec:actual_training}

Many codecs have semantically initialized codebooks 
\cite{that_one_google_paper, reference_primo, llm_codec}, often leveraging a pretrained autoencoder to further compress self-supervised representations \cite{focalcodec, repcodec}. Meanwhile, also common is to attach SD objectives to codebooks in settings where all codec weights are jointly trained \cite{speechtokenizer, xcodec, moshi}. To account for both approaches, we adopt a three-stage training approach shown in Fig. \ref{fig:codec}. 

In the first stage, the \textit{HuBERT codec} and the \textit{pitch codec} are trained to reconstruct respective inputs acquired from the frozen HuBERT model and the deterministic YAAPT module. In addition, the VQ codebook is trained with a commitment loss, defined as the MSE loss between VQ inputs and outputs. Following \cite{reference_primo}, exponential moving average updates are used for all VQ codebook entries, while embeddings unused in each batch are randomly reinitialized.

In the second stage, the modified HiFi-GAN vocoder is trained to reconstruct speech while the encoders and VQs are frozen. The training objective is a combination of log-Mel spectrogram L1 loss, adversarial loss, and feature matching loss. The latter two are derived using multi-scale and multi-period discriminators, following \cite{hifigan}. The stage-two reference codec is our baseline.

Finally, in the third stage, the codec is finetuned on three different settings. The HuBERT codec is unfrozen in the third stage, while the pitch codec stays frozen. The training objective is the sum of the second-stage training objective, commitment loss, and one of $\mathcal{L}_{\text{ASR}}$, $\mathcal{L}_{\text{TTR}}$, or a semantic distillation loss ($\mathcal{L}_{\text{HuBERT}}$) defined as the MSE between the input and output of the HuBERT codec. The latter loss is reused from the first stage to simulate past works that use SD objectives on one or more codebooks \cite{speechtokenizer, moshi}. 
For the HuBERT VQ codebook, the first-stage objectives and update schemes are reused.




For all stages, the AdamW optimizer was used with a learning rate of $2\times 10^{-4}$, weight decay of 0.01, and $(\beta_1,\beta_2)=(0.8,0.99)$. Learning rate was reduced every epoch by a factor of $\lambda=0.999$. Validation was run every 1k steps, and training was stopped if validation metrics did not improve for 100k steps.

\subsection{Evaluation}

We conduct two different subjective tests to evaluate the codecs' perceptual quality as well as the intelligibility improvement we claim to achieve using the LM losses. 
First, a MUSHRA-like \cite{mushra-itu-standard} subjective study was conducted to gauge the overall similarity of each codecs' inputs and outputs. 
We used the decoded utterances at 187.5bps for four codec variants: one from the second stage (as explained in Section \ref{sec:reference_codec}), and three from the third stage. 11 audio experts were asked to score 10 sets of utterances on how similar they felt the decoded utterances are compared to the reference. A 3.5kHz LPF low-anchor was used. Second, a mean opinion score (MOS) subjective study evaluates how well the decoded utterances comply to the LJSpeech transcriptions. 18 English speakers were asked to score 15 sets of utterances on how well the utterances match the text, from 1 (``no matching words'') to 7 (``perfectly matches the text''). No low anchor was used. We employed webMUSHRA \cite{webmushra} for both studies, randomly sampling utterances from the test set and normalizing to -24 LUFS.
WARPQ \cite{warpq}, PESQ \cite{pesq}, and WER were also used for acoustic quality and intelligibility. WARPQ and PESQ do not accurately reflect audio quality if temporal alignment lacks, but may be useful for comparisons.
For WER, \texttt{Whisper-large-v3} and \texttt{wav2vec2.0} \cite{wav2vec2} were both used, lest \texttt{Whisper-tiny} in the ASR loss cause unjustly lower WER on the same family of models.




%
\section{Experimental Results}


The MOS results on the right side of Fig. \ref{fig:subjective_results} show that models trained with LM losses boost the semantic compliance of decoded utterances with fewer PHs. In the figure and hereafter, ASR, TTR, and SD denote codecs from stage 3, respectively trained using the proposed ASR loss, TTR loss, and $\mathcal{L}_{\text{HuBERT}}$. S2 denotes the stage-two codec trained on frozen encoder and VQ.
A Wilcoxon signed-rank test with Bonferroni correction shows that all differences except S2 and SD are significant($p<0.05$), meaning that from an experienced English speaker's perspective, the ASR loss-aided codec is the best at preserving phonetic and linguistic content, followed by TTR, then S2 and SD. This result is also validated by Table \ref{tab:the_table}, where ASR and TTR exhibits consistently lower WERs on both ASR models.

Meanwhile, the MUSHRA-style similarity scores from the left side of Fig. \ref{fig:subjective_results} show that ASR, TTR, and SD show comparable performances, while the differences between this group of three and S2 are quite significant. Clearly, both LM losses and SD objective boost overall quality aspects. In summary, our LM losses do not improve overall quality compared to SD, but they significantly outperform S2 (i.e. sans LM losses) in both tests and SD in the semantic test. 

Perhaps most significantly, LM losses seem to widen the breadth of the semantic-acoustic tradeoff \cite{speechtokenizer} than allowed by $\mathcal{L}_{\text{HuBERT}}$. HuBERT-feature-matching losses like $\mathcal{L}_{\text{HuBERT}}$ are known to well-preserve semantic content \cite{repcodec}. Therefore, the second-stage tokens, trained solely using this signal, are an approximate ``upper bound'' as to how much semantic content can be preserved using $\mathcal{L}_{\text{HuBERT}}$, especially since the acoustic losses seem to encourage semantic deviations \cite{speechtokenizer, xy_tokenizer, moshi}.
However, models trained using LM losses clearly surpass the second-stage codecs in both WER and MOS, suggesting that more semantic context has been learned during the third-stage and was stored in the resulting tokens. 

Another reason for the success of LM losses may be that they are end-to-end losses, in contrast to common SD objectives that cannot influence the decoders in any way. Instead of only disentangling away redundant acoustic information in the semantic codes \cite{speechtokenizer}, our decoder is tamed directly by the LM losses as well, so that it may use its generative capability in more linguistically-plausible ways. 

\section{Conclusion}
\label{sec:conclusion}

This work proposed two LM losses and showed that they alleviate PHs in a very-low-bit setting that already includes semantic quality considerations. Both of our proposed ASR and TTR losses seemed to expand the breadth of semantic-acoustic tradeoff than allowed by the SD and acoustic objectives, by boosting the semantic quality of the output while preserving the overall perceived output quality. 
As the LM losses are end-to-end, they can be flexibly applied to any DNN-based codec that outputs speech. The proposed LM losses may prove to be useful for training very-low-bitrate speech codecs when strong semantic complicance is desired.




\vfill\pagebreak





\bibliographystyle{IEEEbib}
\bibliography{refs}

\end{document}